%% file: main_v1_CHI_Workshop.tex
\documentclass[manuscript,screen,nonacm]{acmart}

\usepackage{subcaption}
\usepackage{multirow}
\usepackage{tabularx}
\usepackage{gensymb}
\usepackage{soul}
\usepackage[commandnameprefix=ifneeded]{changes}
\setdeletedmarkup{}
\definechangesauthor[name={Per cusse}, color=blue]{per}
\setauthormarkup{\textcolor{blue}{#1}}

\usepackage{todonotes}
\let\xtodo\todo
\renewcommand{\todo}[1]{\xtodo[inline,color=green!50]{#1}}

\AtBeginDocument{%
  }

\setcopyright{acmcopyright}
\copyrightyear{2018}
\acmYear{2018}
\acmDOI{XXXXXXX.XXXXXXX}
\acmConference[Conference acronym 'XX]{Make sure to enter the correct
  conference title from your rights confirmation email}{June 03--05,
  2018}{Woodstock, NY}
\acmISBN{978-1-4503-XXXX-X/18/06}
\begin{document}

\title{An LLM-Assisted Toolkit for Inspectable Multimodal Emotion Data Annotation}

\author{Zheyuan Kuang}
\orcid{0009-0009-0184-6159}
\affiliation{%
 % \department{School of Computer Science}
 \institution{The University of Sydney}
 \streetaddress{Camperdown/Darlington}
 \city{Sydney}
 % \state{New South Wales}
 \country{Australia}}
\email{zheyuan.kuang@sydney.edu.au}

\author{Weiwei Jiang}
\orcid{0000-0003-4413-2497}
\affiliation{%
 % \department{School of Computer Science}
 \institution{Nanjing University of Information Science and Technology}
 \city{Nanjing}
 % \state{Jiangsu}
 \country{China}}
\email{weiweijiangcn@gmail.com}

\author{Nicholas Koemel}
\orcid{0000-0002-5463-6894}
\affiliation{
 \institution{Charles Perkins Centre, Faculty of Medicine and Health}
 \city{Sydney}
 \country{Australia}}
\affiliation{
 \institution{Turner Institute for Brain and Mental Health, Monash University}
 \city{Melbourne}
 \country{Australia}}
\email{nicholas.koemel@sydney.edu.au}

\author{Matthew Ahmadi}
\orcid{0000-0002-3115-338X}
\affiliation{
 \institution{Charles Perkins Centre, Faculty of Medicine and Health}
 \city{Sydney}
 \country{Australia}}
\affiliation{
 \institution{Turner Institute for Brain and Mental Health, Monash University}
 \city{Melbourne}
 \country{Australia}}
\email{matthew.ahmadi@sydney.edu.au}

\author{Emmanuel Stamatakis}
\orcid{0000-0001-7323-3225}
\affiliation{
 \institution{Charles Perkins Centre, Faculty of Medicine and Health}
 \city{Sydney}
 \country{Australia}}
\affiliation{
 \institution{Turner Institute for Brain and Mental Health, Monash University}
 \city{Melbourne}
 \country{Australia}}
\email{emmanuel.stamatakis@sydney.edu.au}

\author{Benjamin Tag}
\orcid{0000-0002-7831-2632}
\affiliation{%
 % \department{School of Computer Science and Engineering}
 \institution{University of New South Wales}
 \streetaddress{Kensington}
 \city{Sydney}
 % \state{New South Wales}
 \country{Australia}}
\email{benjamin.tag@unsw.edu.au}

\author{Anusha Withana}
\orcid{0000-0001-6587-1278}
\affiliation{%
 % \department{School of Computer Science}
 \institution{The University of Sydney}
 \streetaddress{Camperdown/Darlington}
 \city{Sydney}
 % \state{New South Wales}
 \country{Australia}}
\email{anusha.withana@sydney.edu.au}

\author{Zhanna Sarsenbayeva}
\orcid{0000-0002-1247-6036}
\affiliation{%
 % \department{School of Computer Science}
 \institution{The University of Sydney}
 \streetaddress{Camperdown/Darlington}
 \city{Sydney}
 % \state{New South Wales}
 \country{Australia}}
% \email{zsar5653@sydney.edu.au}
\email{zhanna.sarsenbayeva@sydney.edu.au}

\renewcommand{\shortauthors}{Kuang et al.}

\begin{abstract}
Multimodal Emotion Recognition (MER) increasingly depends on fine grained, evidence grounded annotations, yet inspection and label construction are hard to scale when cues are dynamic and misaligned across modalities. We present an LLM-assisted toolkit that supports multimodal emotion data annotation through an inspectable, event centered workflow. The toolkit preprocesses and aligns heterogeneous recordings, visualizes all modalities on an interactive shared timeline, and renders structured signals as video tracks for cross modal consistency checks. It then detects candidate events and packages synchronized keyframes and time windows as event packets with traceable pointers to the source data. Finally, the toolkit integrates an LLM with modality specific tools and prompt templates to draft structured annotations for analyst verification and editing. We demonstrate the workflow on multimodal VR emotion recordings with representative examples.
\end{abstract}

%% The code below is generated by the tool at http://dl.acm.org/ccs.cfm.


% \ccsdesc[500]{Computing methodologies~Machine learning}


%% A "teaser" image appears between the author and affiliation
%% information and the body of the document, and typically spans the
%% page.
% \begin{teaserfigure}
%  \centerline{\includegraphics[width=\textwidth]{Figure/teaser.pdf}}
%  \caption{Illustration of the participant under six virtual reality scenes that could elicit emotions with interactive objects highlighted in color. }
%  \Description{}
%  \label{fig:teaser}
% \end{teaserfigure}

\maketitle

\input{Section/1-Introduction}

\input{Section/2-RelatedWork}

\input{Section/3-ProposeApproach}

\input{Section/5-ConclusionAndFuturework}

% \begin{acks}
% To Robert, for the bagels and explaining CMYK and color spaces.
% \end{acks}

\bibliographystyle{ACM-Reference-Format}
\bibliography{bibliography}

\end{document}

%% file: Section/1-Introduction.tex
\section{Introduction}

Multimodal Emotion Recognition (MER) faces practical challenges that often stem from data work rather than model design. Emotion related cues are distributed across modalities and vary over time, making cross modal integration and event level interpretation difficult. This increases the demand for fine grained, descriptive, evidence grounded annotations. However, benchmark results highlight that annotation cost, limited labeled data, and reduced robustness under domain gaps, noise, and missing modalities remain major barriers~\cite{lian2025mer,lian2026merbench}. These constraints make scalable inspection and consistent label construction a central challenge for MER.

Building on these challenges, recent MER work has started to leverage LLMs to support more descriptive and open ended emotion understanding~\cite{lian2025mer}. In practice, a limitation in many MER studies is data inspection, where analysts manually align multiple videos and signal plots to track dynamic, often misaligned cues across modalities and to decide which time windows should be treated as candidate events~\cite{puccetti2022temporal}. Within this data work context, LLMs offer a complementary capability by generating structured multimodal annotations in a consistent format, reducing manual annotation effort while supporting analyst verification and editing~\cite{zhang2023large, wang2024human}.

We address these gaps by developing an LLM assisted toolkit\footnote{The toolkit repository will be released as open source upon acceptance of the paper.} that operationalizes a three stage workflow for multimodal emotion data annotation. The toolkit first preprocesses and aligns heterogeneous recordings, then visualizes all modalities on an interactive shared timeline, rendering structured signals as video tracks to support inspectable cross modal consistency checks. Second, it detects candidate events and retrieves corresponding keyframes and time windows as event packets with traceable pointers. Third, it uses an LLM with modality specific tools and prompt templates to draft structured annotations that analysts can verify and edit. Together, these stages support scalable construction of training ready labels from multimodal emotion datasets. 
We demonstrate the workflow on multimodal VR emotion recordings with representative examples.

%% file: Section/2-RelatedWork.tex
\section{Related Work}

Recent work has begun to treat LLMs as part of the data work workflow, where LLMs collaborate on preprocessing and inspection, but still require human verification and explicit records of transformations~\cite{kang2024human}. Researchers explore LLMs as data preprocessors for tasks such as error detection, imputation, and matching, and highlight practical limitations that motivate careful checking~\cite{zhang2023large}. Visual analytics research similarly argues that LLMs can help people navigate, query, and summarize complex data during interactive inspection, while stressing grounding and risk management rather than blind automation~\cite{hutchinson2024llm}.  In annotation, Human–LLM collaborative workflows show that LLMs can propose labels and explanations while humans validate uncertain cases, improving efficiency without removing human control~\cite{wang2024human}. 

LLMs also support emotion dataset construction by generating candidate labels, with quality managed through human checks or model validation. For example, \citet{jing2025melt} apply GPT-4o with structured prompting to annotate a multimodal emotion dataset, and validate labels through human review and downstream training, showing both utility and the need for verification. \citet{niu2025rethinking} show that LLM emotion judgments can systematically differ from human annotations, and argue for integrating LLMs mainly for triage and quality control rather than fully replacing humans.

%% file: Section/3-ProposeApproach.tex
\section{Proposed Approach}

We present an LLM assisted toolkit that operationalizes a three stage data handling workflow for multimodal emotion data annotation. The toolkit transforms heterogeneous recordings into inspectable, event-centered outputs that support visual inspection of signal changes, emotion-related event detection and keyframe retrieval, and evidence-grounded annotations. It supports data modalities including but not limited to video streams and structured time-series signals. 

\begin{figure}[t]
  \centering
  \includegraphics[width=\linewidth]{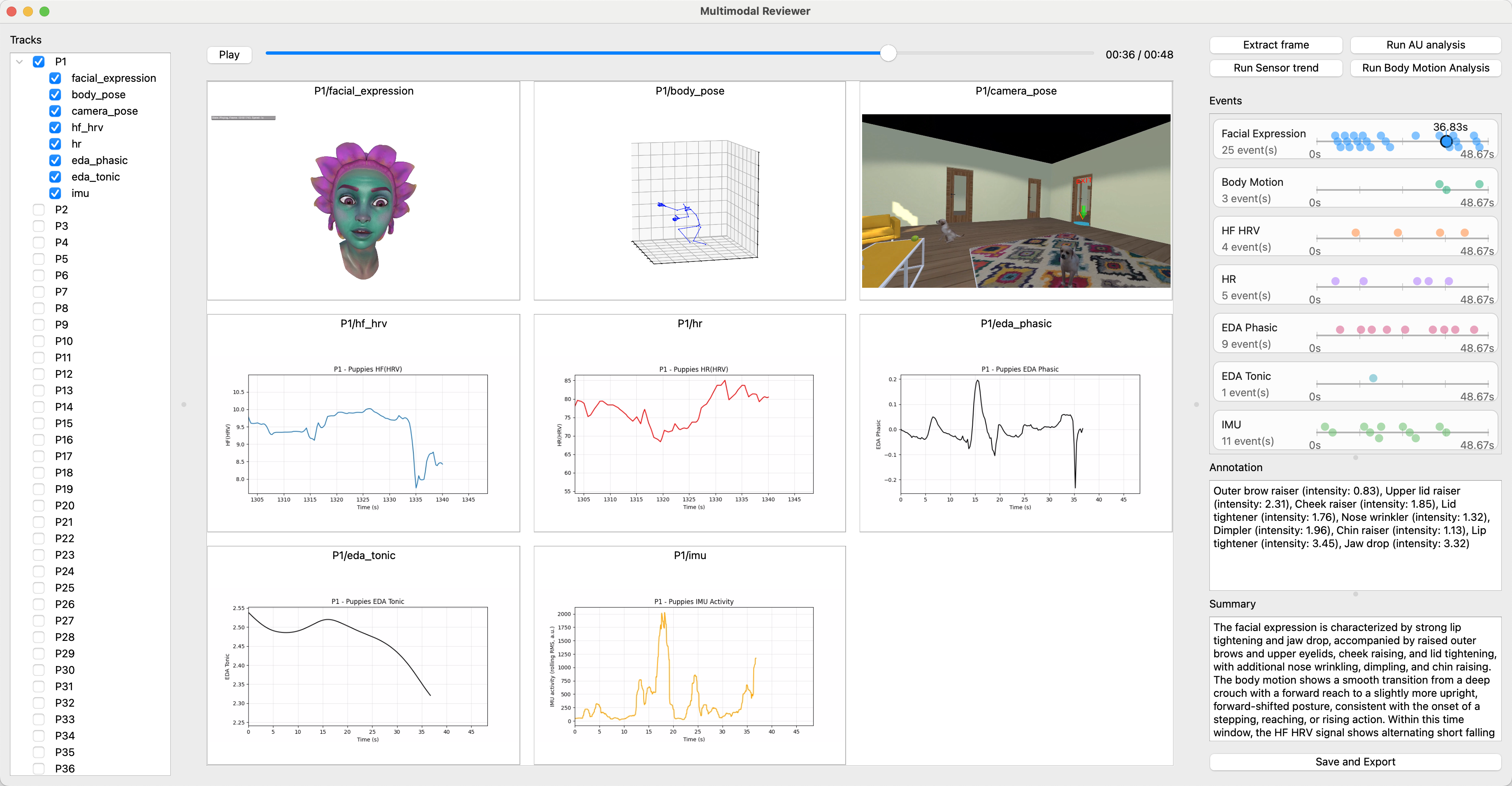}
  \caption{Toolkit interface for multimodal emotion data inspection and annotation.}
  \Description{UI}
  \label{fig:ui}
\end{figure}

We demonstrate the toolkit on our multimodal VR emotion recordings from 84 participants across six emotion elicitation scenes, including avatar-based facial expression video, first-person view video, structured body motion, blood volume pulse (BVP), heart rate (HR), electrodermal activity (EDA), and inertial measurement unit (IMU) signals. As shown in~\autoref{fig:ui}, the interface provides synchronized multimodal tracks, event timelines, and annotation views for event-centered inspection and labeling.

\subsection{Preprocess and visualization}

In the first stage, the toolkit performs preprocessing and unified visualization. The goal is to make heterogeneous modalities jointly inspectable on a shared timeline so that analysts can efficiently observe signal changes and localize potential events. We first standardize each modality with consistent session metadata, including participant identifiers, VR scene identifiers, sampling rates, and timestamps. For structured signals, we render each stream into a video track, so that time series can be inspected with the same interaction paradigm as videos. This enables time locked playback and side by side comparison across different modality tracks in a single interface. 

We then synchronize all tracks to a unified time axis and persist the alignment as an index, which supports navigation, zooming, and reproducible retrieval of any time window across modalities. The resulting UI organizes sessions by participant and VR scene, providing an overview for rapid scanning and detailed views for inspecting fine grained temporal patterns behavioral and physiological signal peaks. This stage outputs aligned visual tracks and a session level index that serve as the basis for event mining and subsequent annotation.
 
\subsection{Event detection and retrieval}

In the second stage, the toolkit detects signal change events and retrieves the corresponding keyframes and time windows across modalities. For facial expression videos, we perform Action Units (AUs)~\cite{ekman1978facial} peak detection using OpenFace~\cite{baltruvsaitis2016openface}, and treat each peak as a candidate event. For body motion signals, we compute motion energy measures to identify motion events~\cite{patrona2018motion}. For physiological signals, we extract peak and trend change windows to capture candidate events. All candidates are aligned to the unified time axis and aggregated into an event layer in the UI, enabling analysts to jump to events, compare cross-modal context, and refine event boundaries. 
We render this layer as an interactive event timeline, where clicking an event marker seeks to its timestamp and shows the aligned frames across modalities.
Analysts can verify, edit, or discard candidate events in the UI. The toolkit then packages each event into an event packet with traceable pointers to the original data, including keyframes and time windows for all modalities.

\subsection{LLM annotations}

In the third stage, the toolkit supports LLM assisted emotional annotation for extracted events. The goal of this stage is to convert each event window into standardized annotations and evidence summaries that can be used for emotion related event detection and for constructing training data for downstream emotion recognition models.

We designed a set of modality specific preprocessing tools and prompt templates to guide the LLM to produce a structured annotation with consistent fields. 
We employ GPT-5.2 as the backend LLM to leverage its multimodal reasoning capabilities.
For facial expression, we map peak AU frames to short textual descriptions~\cite{cheng2024emotion}. For body motion, the LLM generates annotations that describe the skeleton sequence within the event window using concise posture and movement cues~\cite{lu2025understanding}. For physiological signals, the LLM summarizes the window using trend and peak descriptors, including rising or falling segments, local extrema, and duration~\cite{li2025sensorllm}. For the first person view stream, the LLM captures contextual information such as activities and environment cues that support interpretation of the event. Finally, the LLM refines the annotations by aggregating unimodal descriptions into a detailed multimodal description for each event~\cite{cheng2024emotion}, which is added to the event packet as an additional emotional descriptor.

Analysts review the generated annotation in the UI, jump to the referenced time ranges for verification, and edit or discard incorrect fields. Finalized annotations are then exported as structured records for downstream use.

%% file: Section/5-ConclusionAndFuturework.tex
\section{Conclusion and Future Work}

In this paper, we present an LLM-assisted toolkit for inspectable, event-centered multimodal emotion data annotation. The toolkit preprocesses and aligns heterogeneous recordings on an interactive shared timeline, renders structured signals as video tracks, and supports candidate event retrieval. For each event, an LLM generates annotations that analysts can verify and edit, supporting more scalable and consistent MER data work.

In future work, we will evaluate the toolkit through a user study with domain experts, comparing inspection efficiency and annotation outcomes against standard practices. We also plan to validate the workflow on additional datasets and extend the toolkit to additional modalities and more adaptable annotation schemas.

%% file: bibliography.bib
@String{Computing = "Computing" }

@article{cheng2024emotion,
  title={Emotion-llama: Multimodal emotion recognition and reasoning with instruction tuning},
  author={Cheng, Zebang and Cheng, Zhi-Qi and He, Jun-Yan and Wang, Kai and Lin, Yuxiang and Lian, Zheng and Peng, Xiaojiang and Hauptmann, Alexander},
  journal={Advances in Neural Information Processing Systems},
  volume={37},
  pages={110805--110853},
  year={2024}
}

@inproceedings{li2025sensorllm,
  title={Sensorllm: Aligning large language models with motion sensors for human activity recognition},
  author={Li, Zechen and Deldari, Shohreh and Chen, Linyao and Xue, Hao and Salim, Flora D},
  booktitle={Proceedings of the 2025 Conference on Empirical Methods in Natural Language Processing},
  pages={354--379},
  year={2025}
}

@article{zhang2023large,
  title={Large language models as data preprocessors},
  author={Zhang, Haochen and Dong, Yuyang and Xiao, Chuan and Oyamada, Masafumi},
  journal={arXiv:2308.16361},
  year={2023}
}

@article{hutchinson2024llm,
  title={LLM-assisted visual analytics: Opportunities and challenges},
  author={Hutchinson, Maeve and Jianu, Radu and Slingsby, Aidan and Madhyastha, Pranava},
  journal={arXiv:2409.02691},
  year={2024}
}

@inproceedings{wang2024human,
  title={Human-llm collaborative annotation through effective verification of llm labels},
  author={Wang, Xinru and Kim, Hannah and Rahman, Sajjadur and Mitra, Kushan and Miao, Zhengjie},
  booktitle={Proceedings of the 2024 CHI Conference on Human Factors in Computing Systems},
  pages={1--21},
  year={2024}
}

@inproceedings{kang2024human,
  title={Human-in-the-loop synthetic text data inspection with provenance tracking},
  author={Kang, Hong Jin and Gulzar, Muhammad Ali and Peng, Nanyun and Kim, Miryung and others},
  booktitle={Findings of the Association for Computational Linguistics: NAACL 2024},
  pages={3118--3129},
  year={2024}
}

@article{jing2025melt,
  title={MELT: Towards Automated Multimodal Emotion Data Annotation by Leveraging LLM Embedded Knowledge},
  author={Jing, Xin and Wang, Jiadong and Tsangko, Iosif and Triantafyllopoulos, Andreas and Schuller, Bj{\"o}rn W},
  journal={arXiv:2505.24493},
  year={2025}
}

@article{niu2025rethinking,
  title={Rethinking emotion annotations in the era of large language models},
  author={Niu, Minxue and El-Tawil, Yara and Romana, Amrit and Provost, Emily Mower},
  journal={IEEE Transactions on Affective Computing},
  year={2025},
  publisher={IEEE}
}

@article{ekman1978facial,
  title={Facial action coding system},
  author={Ekman, Paul and Friesen, Wallace V},
  journal={Environmental Psychology \& Nonverbal Behavior},
  year={1978}
}

@inproceedings{baltruvsaitis2016openface,
  title={Openface: an open source facial behavior analysis toolkit},
  author={Baltru{\v{s}}aitis, Tadas and Robinson, Peter and Morency, Louis-Philippe},
  booktitle={WACV},
  pages={1--10},
  year={2016}
}

@article{patrona2018motion,
  title={Motion analysis: Action detection, recognition and evaluation based on motion capture data},
  author={Patrona, Fotini and Chatzitofis, Anargyros and Zarpalas, Dimitrios and Daras, Petros},
  journal={Pattern Recognition},
  volume={76},
  pages={612--622},
  year={2018},
  publisher={Elsevier}
}

@inproceedings{lu2025understanding,
  title={Understanding emotional body expressions via large language models},
  author={Lu, Haifeng and Chen, Jiuyi and Liang, Feng and Tan, Mingkui and Zeng, Runhao and Hu, Xiping},
  booktitle={Proceedings of the AAAI Conference on Artificial Intelligence},
  volume={39},
  number={2},
  pages={1447--1455},
  year={2025}
}

@inproceedings{lian2025mer,
  title={Mer 2025: When affective computing meets large language models},
  author={Lian, Zheng and Liu, Rui and Xu, Kele and Liu, Bin and Liu, Xuefei and Zhang, Yazhou and Liu, Xin and Li, Yong and Cheng, Zebang and Zuo, Haolin and others},
  booktitle={Proceedings of the 33rd ACM International Conference on Multimedia},
  pages={13837--13842},
  year={2025}
}

@article{lian2026merbench,
  title={Merbench: A unified evaluation benchmark for multimodal emotion recognition},
  author={Lian, Zheng and Sun, Licai and Ren, Yong and Gu, Hao and Sun, Haiyang and Chen, Lan and Liu, Bin and Tao, Jianhua},
  journal={IEEE Transactions on Pattern Analysis and Machine Intelligence},
  year={2026},
  publisher={IEEE}
}

@article{puccetti2022temporal,
  title={Temporal dynamics of affect in the brain: Evidence from human imaging and animal models},
  author={Puccetti, Nikki A and Villano, William J and Fadok, Jonathan P and Heller, Aaron S},
  journal={Neuroscience \& Biobehavioral Reviews},
  volume={133},
  pages={104491},
  year={2022},
  publisher={Elsevier}
}
